# Why Do You Spread This Message? Understanding Users Sentiment in Social Media Campaigns

**Jalal Mahmud[1] and Huiji Gao[2]**

[1]IBM Research - Almaden
jumahmud@us.ibm.com

[2]Arizona State University
Huiji.Gao@asu.edu

### Abstract

Twitter has been increasingly used for spreading messages about campaigns. Such campaigns try to gain followers through their Twitter accounts, influence the followers and spread messages through them. In this paper, we explore the relationship between followers' sentiment towards the campaign topic and their rate of retweeting of messages generated by the campaign. Our analysis with followers of multiple social-media campaigns found statistical significant correlations between such sentiment and retweeting rate. Based on our analysis, we have conducted an online intervention study among the followers of different social-media campaigns. Our study shows that targeting followers based on their sentiment towards the campaign can give higher retweet rate than a number of other baseline approaches.

## Introduction

Recent years have seen a rapid growth in micro-blogging and the rise of popular micro-blogging services such as Twitter. With the growing usage of such micro-blogging services, a wide number of social-media campaigns ranging from politics and government to social issues also exist in Twitter. Such campaigns maintain Twitter accounts, want to gain a large number of followers through their Twitter accounts, influence the followers and spread messages through them. However, in reality, all followers may not be equally engaged with the campaign (Chen et al. 2012). Chen et al. found that users' activity level, prior topic interest, prior interpersonal relation and geographical location is correlated with their level of engagement (measured by retweets and hashtag usage) with Occupy Wall Street[1] campaign (Chen et al. 2012).

Motivated by their findings, we explore whether followers' sentiment towards the campaign topic is correlated with their engagement level with the campaign, where engagement is measured as the rate of retweeting of the messages generated by the campaign. Furthermore, we investigate whether such retweet rate can be predicted from their sentiment towards the campaign topic.

Our research is also inspired by an established theoretical framework in psychological and marketing research on *attitudes* and *attitude models*, where attitude is defined as a unified concept containing three aspects: "*feelings*", "*beliefs*", and "*actions*" (Schiffman et al. 2010). According to the framework, beliefs are acquired on *attitude object* (e.g., a topic or product), which in turns influences the feelings on the object and the actions with regard to the attitude object. Since user's belief is hard to observe from social media data, we focused on understanding relation between feelings (sentiment) towards a campaign topic and actions which result from such feelings (sentiment). We hypothesize that user's sentiment towards social media campaign has positive effects on their actions (e.g., retweet) related to the campaign, and validate this hypothesis by this research. We have also conducted an online intervention study to understand the effect of sentiment on increasing retweet rates of campaign relevant messages.

Our study shows that targeting followers of a campaign based on their higher positive sentiment towards the campaign gives higher retweeting rate than a number of baselines, such as random targeting and topic based targeting. Thus, a campaign can be more effective by sending targeted messages to followers with stronger sentiment towards the campaign topic.

## Dataset

For the sake of concreteness, we limit our exploration on the campaign topic "fracking"[2], and analyze data from Twitter accounts of several campaigns either supporting or opposing "fracking". We manually selected 4 Twitter campaigns related to "fracking" on July 31, 2013. Table 1 shows the data-statistics of the campaigns. We crawled their tweets from last one month and identified tweets which contained keywords or hashtags related to fracking, such as "fracking", "shale", "#fracking", "#shale", "oil". We denoted these tweets as *topic relevant* tweets. We also manually inspected those tweets and found that they were indeed topically relevant. Table 2 shows few such tweets with their sentiment. The rest of the tweets of each campaign in last one month are denoted as *general* tweets. Then, we crawled followers of each campaign and their tweets from last one month. These tweets are denoted as *recent* tweets. However, we excluded followers who had protected accounts or who had posted fewer than 100 tweets. We also crawled historical tweets (200 max.) of those followers before last one month (tweets until July 1, 2013), and denoted such tweets as *historical* tweets.

---

[1] https://twitter.com/OccupyWallSt

[2] fracking or hydrolic fracturing is the process of extracting natural gas from shale rock layers. This is very controversial due to its potential impact on energy and environment

| Twitter account name of the campaign | Type (pro/anti fracking) | # of followers as of July 31, 2013 | # of Tweets posted as of July 31, 2013 |
|---|---|---|---|
| shalebiz | pro | 2776 | 6940 |
| Energy From Shale | pro | 3965 | 4826 |
| Fracking News | anti | 1219 | 3062 |
| Frackfree America | anti | 1042 | 6133 |

**Table 1. Social Media Campaigns in our dataset**

| Sentiment | Tweet |
|---|---|
| Positive | The science & economics of #fracking says yes to fracking http://ow.ly/ll1J4 |
| Negative | #Fracking wastewater threatens to drown Ohio: http://t.co/ft768gkW |

**Table 2. Example tweets related to fracking**

## Sentiment and Retweet Behavior

Our goal is to investigate whether followers' sentiment towards the campaign topic has any correlation with how they engage with the campaign. However, we do not have ground truth sentiment for such followers. Hence, we inferred their sentiment from historical tweets.

### Sentiment Prediction Model

Motivated by prior works on sentiment analysis (Barbosa et al. 2010, Davidov et al. 2010), we developed a simple content-based sentiment prediction model which predicts the sentiment of a Twitter user towards "fracking". We first created a labeled dataset for sentiment prediction from Twitter. We used Twitter's streaming API from January, 2013 to March, 2013 for tweets containing keywords or hashtags related to fracking (e.g., "fracking", "shale", "#fracking", "#shale", "oil"). In total, we collected about 1.68 million tweets. We identified retweets from our data, and computed how many times each tweet was retweeted.

We selected the tweets which were retweeted at least 100 times, and manually analyzed them for ground-truth creation. In particular, we found 163 such tweets which we labeled as either positive or negative towards "fracking". 22 tweets were positive (pro-fracking tweets), and 141 tweets were negative (anti-fracking tweets). From our data, we identified all users who had retweeted those tweets. There were 5384 such users in total: 1562 users retweeted pro-fracking tweets, and 3822 users retweeted anti-fracking tweets. We considered the users who retweeted pro-fracking tweets as positive towards "fracking", and those who retweeted anti-fracking tweets as negative towards "fracking". This is based on the traditional assumption that retweet is an act of endorsement of the original tweet (Boyd et al. 2010). From the 5384 users (1562 positive, 3822 negative), we randomly sampled 1000 positive (pro-fracking) users and another 1000 negative (anti-fracking) users. Then, we used Twitter's REST API to crawl the historical tweets (200 max.) of those users before the retweet. Furthermore, we used Twitter's streaming API to randomly sample another 1000 users who were assumed to have neutral sentiment towards "fracking". We also crawled their historical (200 max.) tweets. Thus, our final dataset for sentiment prediction model contained 3000 users: 1000 positive, 1000 negative and 1000 neutral towards "fracking".

From the above data, we constructed classification-based model with three categories (positive, negative, neutral). These are *sentiment polarity* for users' sentiment prediction. Our model used unigrams computed from historical tweets as features. We tried a number of different statistical models from WEKA, a widely used machine learning toolkit. SVM-based models outperformed others, and achieved ~92% accuracy (0.92 recall, 0.93 precision, 0.925 F-measure) under 10-fold cross-validation.

### Accuracy of Sentiment Prediction for Campaign Followers

We conducted a study where participants labeled the sentiment (positive, negative, neutral) of the followers by looking at their historical tweets (tweets until July 1, 2013). Then, we compared this manually labeled sentiment with inferred sentiment by our algorithm and computed accuracy. We randomly selected 50 followers from each campaign (200 followers in total) and collected their historical tweets. We recruited 100 participants from CrowdFlower (http://crowdflower.com/), a popular crowd sourcing platform. Each of them labeled 4 followers, and each of the follower was labeled by two participants. Participants were in agreement for 126 followers, and we used those as ground-truth to compute the accuracy of our sentiment prediction model. We applied our sentiment prediction model to infer the sentiment of those 126 followers. Our model's inferred sentiment matched manually labeled sentiment for 110 users which corresponds to 87.3% accuracy.

| Twitter account name of the campaign | Correlation Analysis | | | |
|---|---|---|---|---|
| | normalized-retweet-rate-of-topic-related-campaign-msg | | normalized-retweet-rate-of-general-campaign-msg | |
| | Corr. with Sent. polarity | Corr. with Sent. strength | Corr. with Sent. polarity | Corr. with Sent. strength |
| Shalebiz | 0.65* | 0.7* | 0.6 | 0.62* |
| Energy From Shale | 0.71* | 0.73* | 0.64 | 0.65* |
| Fracking News | 0.8* | 0.82* | 0.77 | 0.8* |
| Frackfree America | 0.6* | 0.63* | 0.5 | 0.55* |

**Table 3. Pearson correlation between sentiment and normalized retweet rates (* means significant, p < 0.05)**

| Twitter account name of the campaign | Mean Absolute Error (MAE) | | | |
|---|---|---|---|---|
| | Predicting topical msg | | Predicting general msg | |
| | Pred. from Sent. polarity | Pred. from Sent. strength | Pred. from Sent. polarity | Pred. from Sent. strength |
| shalebiz | 0.32 | 0.25 | 0.35 | 0.26 |
| Energy From Shale | 0.30 | 0.24 | 0.32 | 0.28 |
| Fracking News | 0.28 | 0.25 | 0.33 | 0.27 |
| Frackfree America | 0.27 | 0.23 | 0.3 | 0.25 |

**Table 4. Regression results over 10 fold cross-validation**

## Correlation of Sentiment with Retweets

We applied the sentiment prediction model to infer the sentiment of followers of each campaign from their historical tweets (tweets until July 1, 2013). Then, we investigated whether followers' predicted sentiment towards fracking has any correlation with their actions (e.g., retweets). Towards that, we analyzed the *recent* tweets of each follower, and computed the number of times they retweeted any *topic relevant* tweets of the campaign. We divided that number by the total number of topic relevant tweets of the campaign. The resultant ratio is named as *retweet-rate-of-topic-related-campaign-msg*.

We do a similar analysis for other tweets generated from the campaign and computed the ratio: *retweet-rate-of-general-campaign-msg*. Since some followers can be more active in retweeing than others, we divided each of these ratios by the total number of retweets of each follower over last one month. This resulted normalized retweet rates which we used for our analysis.

Next, we conducted a pearson correlation analysis between sentiment and retweet rates. For such analysis, we converted predicted sentiment to numeric scores. For followers of pro-fracking campaigns, the numeric scores are as follows: 1 for positive, 0 for neutral and -1 for negative. For followers of anti-fracking campaigns, we do the opposite (1 for negative, 0 for neutral and -1 for positive). Our sentiment prediction model also returned a probability estimate associated with sentiment prediction. We multiplied this probability with the numeric score for predicted sentiment to obtain *sentiment strength* of each follower. Then, we also investigated the correlation between sentiment strength and retweet rates.

As shown in Table 3, we found positive correlations between sentiment polarity, and retweet rates for all campaigns. This indicates that followers who express similar sentiment of the campaign are also more likely to retweet messages generated by the campaign. For retweeting *topic relevant* campaign tweets, such correlations are statistically significant for all campaigns. However, for retweeting *general* campaign tweets, correlations are not statistically significant. Table 3 also shows the correlations between sentiment strength and retweeting rates. Statistical significant correlations are observed in each case. We also see a higher correlation value for sentiment strength, which indicates its effectiveness to impact followers' retweets.

## Prediction Model

We also built predictive models of retweeting a campaign's message based on the sentiment of the followers. We performed both regression analysis and a classification study using WEKA, a widely used machine learning toolkit. For both cases, we tried two settings - predicting from sentiment polarity, and predicting from sentiment strength.

For regression analysis, sentiment polarity (first setting) and sentiment strength (second setting) are used as the value of the independent variable in the regression model. Retweet rates are used as dependent variable in the model. We tried a number of regression approaches (e.g., logistic regression, SVM regression) and performed 10-fold cross validation. Logistic regression performed the best, and the results in terms of mean absolute error (MAE) are shown in Table 4. We find lower prediction errors when predicting from sentiment strength. In particular, retweeting topi-cal messages can be predicted within 23%-25% MAE and general messages can be predicted within 25%-28% MAE.

| Twitter account name of the campaign | Area Under ROC Curve (AUC) | | | |
|---|---|---|---|---|
| | Predicting topical msg | | Predicting general msg | |
| | Pred. from Sent. polarity | Pred. from Sent. strength | Pred. from Sent. polarity | Pred. from Sent. strength |
| shalebiz | 0.67 | 0.74 | 0.63 | 0.70 |
| Energy From Shale | 0.69 | 0.76 | 0.67 | 0.72 |
| Fracking News | 0.70 | 0.77 | 0.66 | 0.73 |
| Frackfree America | 0.73 | 0.80 | 0.70 | 0.74 |

**Table 5. Classification results over 10 fold cross-validation**

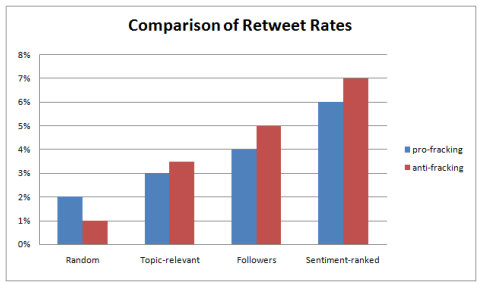

**Figure 1.** Retweet rates for different target groups (500 messages were sent in each case)

For classification study, we used supervised binary machine learning algorithms to classify followers with above median levels of retweet rates. We experimented with a number of classifiers from WEKA, including naive Bayes, SVM, J48, Random Forest. SVM slightly outperformed the rest. Table 5 shows the classification result in terms of AUC under 10-fold cross validation. We see that classifying high (above median) or low (below median) retweeters from sentiment can be done quite accurately.

| Sentiment | Example Intervention Message |
|---|---|
| Positive | "@zas Plz RT "Fracking saves us money; fracking creates jobs"" (pro-fracking) |
| Negative | "To anyone speaking of the economic "benefits" of fracking: what use is that money if your food and water are full of poison. Plz RT"(anti-fracking) |

**Table 6.** Example intervention messages

## Online Engagement Study

We conducted an engagement study among followers with different sentiment. In our study, we sent campaign relevant messages to such followers. We first created 6 accounts on Twitter, 3 for sending pro-fracking and another 3 for sending anti-fracking messages. We maintained a pool of 10 pro-fracking messages and 10 anti-fracking messages. Each such message asked the recipient to retweet the message.

We ranked the positive sentiment followers of pro-fracking campaigns according to their sentiment strength towards the "fracking" topic and selected top-500 (*sentiment-ranked-top-followers*) from the ranked list. From the remaining pro-fracking followers, we randomly selected 500 followers and denoted them as *followers*. We obtained another set of 500 users from random sampling from Twitter stream. Finally, we looked for Twitter stream for users who mentioned the term "fracking" and obtained another set of 500 users from such matched users (*topic-relevant*).

From our pro-fracking accounts, we sent pro-fracking messages to each group of users. Thus, 500 messages were sent for each target group. In each such message sending, we randomly selected a message from the pool of 10 pro-fracking messages. We waited about a week, and recorded how many messages were retweeted in each case. Then, we computed the retweet rates (defined as the ratio of the number of messages retweeted and the total number of messages sent) obtained in each case.

We do a similar study for anti-fracking followers. Thus, we sent anti-fracking messages to *sentiment-ranked-top-followers* of anti-fracking campaigns. We also sent those messages to randomly selected 500 followers of those campaigns, randomly selected 500 users from Twitter, and randomly selected 500 topic-relevant users from Twitter. Figure 1 shows the retweet rates obtained in each case. We observe that sentiment-ranked approach clearly outperformed the other approaches. Our study shows that a campaign can be more effective by sending targeted messages to followers with stronger sentiment towards the campaign.

## Conclusion

In this paper, we investigated the relationship between sentiment of social media users who are following a campaign and their likelihood of spreading campaign messages. We found that sentiment has a statistically significant effect on such activity. Furthermore, spreading topically relevant campaign messages has stronger correlation with sentiment of the campaign followers. Our engagement study provides further insight for designing better intervention to spread campaign messages. There are various directions for future research. First, we like to apply similar analysis for other types of actions such as hashtag usage, mentions or tweet creation. Second, we will explore how other factors (e.g., general activity, prior interaction as suggested by Chen et al. 2012) together with sentiment predict such actions. Third, we will conduct a survey among campaign followers to understand their demographics, personality, network size, etc. and analyze whether these factors affect their engagement with the campaign. Finally, we would like to investigate the generality of our findings by applying similar analysis for social media campaigns on different topics.